\documentclass[aps,pre,twocolumn,showpacs,superscriptaddress,groupedaddress]{revtex4-2}
\usepackage{amsmath}
\usepackage{amssymb}
\usepackage{amsfonts}
\usepackage{graphicx}
\usepackage{dcolumn}
\usepackage{bm}
\usepackage{sidecap}
\usepackage{subfloat}
\usepackage{parskip}
\usepackage{grffile}
\usepackage{color}
\usepackage{hyperref}
\usepackage{hhline}
\usepackage{mathtools}
\usepackage{graphics}
\usepackage{multirow}
\usepackage{verbatim}
\usepackage{longtable}
\usepackage{rotating}
\usepackage{setspace}
\usepackage{epsfig}
\usepackage{subfigure}
\usepackage{epstopdf}
\usepackage{gensymb}
\usepackage[normalem]{ulem}
\makeatletter
\def\@eqnnum{{\normalsize \normalcolor (\theequation)}}
 \makeatother

\graphicspath{{./}{ER/main/}}

\begin{document}

%------------------------------
\title{Explosive synchronization and chimera in interpinned multilayer networks}
\author{Ajay Deep Kachhvah$^1$, and Sarika Jalan$^{1,2}$}
\affiliation{1. Complex Systems Lab, Department of Physics, Indian Institute of Technology Indore, Khandwa Road, Simrol, Indore-453552, India}
\affiliation{2. Department of Biosciences and Biomedical Engineering, Indian Institute of Technology Indore, Khandwa Road, Simrol, Indore-453552, India}

%\date{\today}

\begin{abstract}
This Letter investigates the nature of synchronization in multilayered and multiplexed populations in which the interlayer interactions are randomly pinned. First, we show that a multilayer network constructed by setting up all-to-all interlayer connections between the two populations leads to explosive synchronization in the two populations successively, leading to the coexistence of coherent and incoherent populations forming chimera states. Second, a multiplex formation of the two populations in which only the mirror nodes are interconnected espouses explosive transitions in the two populations concurrently. The occurrence of both explosive synchronization and chimera are substantiated with rigorous theoretical mean-field analysis. The random pinning in the interlayer interactions concerns the practical problems where the impact of dynamics of one network on that of other interconnected networks remains elusive, as is the case for many real-world systems.
\end{abstract}


\maketitle

\paragraph*{\bf{Introduction}}
The dynamical evolution of large-scale complex systems having underlying graph structures has been popularly modeled using coupled Kuramoto oscillators on networks~\cite{Kuramoto1984,Acebron2005}. A multilayer network, which refers to the same sets of nodes having different types of interactions among its units, has brought forward many astonishing phenomena and sheds light on the mechanisms behind emerging behaviors beyond a single layer framework~\cite{Boccaletti2014,Osat2017,Nicosia2017,Pitsik2018,Danziger2019,Rybalova2020,Berner2020,Totz2020,Shepelev2021}. One such behavior is explosive synchronization (ES), which refers to the first-order transition to synchronization~\cite{Tanaka1997,Pomerening2003,Mirollo2005,Gaedenes2011,Leyva2012,Danziger2016,AGaytan2018,Chandrasekar2020,Kuehn}. In contrast to a smooth transition to synchronization, an abrupt jump to the coherence accompanied by hysteresis is witnessed as the strength of couplings between the interacting units increases. 
It is reported that the dynamical or structural features such as delay, phase lag, and weight adaptation employed in the intralayer or interlayer couplings in multilayer networks play a crucial role in controlling the characteristics of emergent ES transition~\cite{Kachhvah2019,anil2021,Kachhvah2020,Zhang2015,Khanra2021,Frolov2021}.
 
Further, a chimera state (CS) refers to the coexistence of coherent and incoherent states, which is an upshot of the partial symmetry breaking of the system~\cite{Abrams2004,Asir2021}. A decade ago, Abrams \textit{et al.}~\cite{Abrams2008} reported the breathing chimera states in two groups of identically coupled phase lagged oscillators. Recently, borrowing the same model but composed of nonidentical oscillators, the study was extended for a wider parameter space exhibiting various chimera states~\cite{Guo2021}. Later, a model considering two groups of the phase lagged nonidentical oscillators in the presence of adaptively controlled coupling reported the bridging of ES with the chimera state~\cite{Zhang2016}. 
\begin{figure}[ht]
	\centering
	\includegraphics[height=4cm,width=8cm]{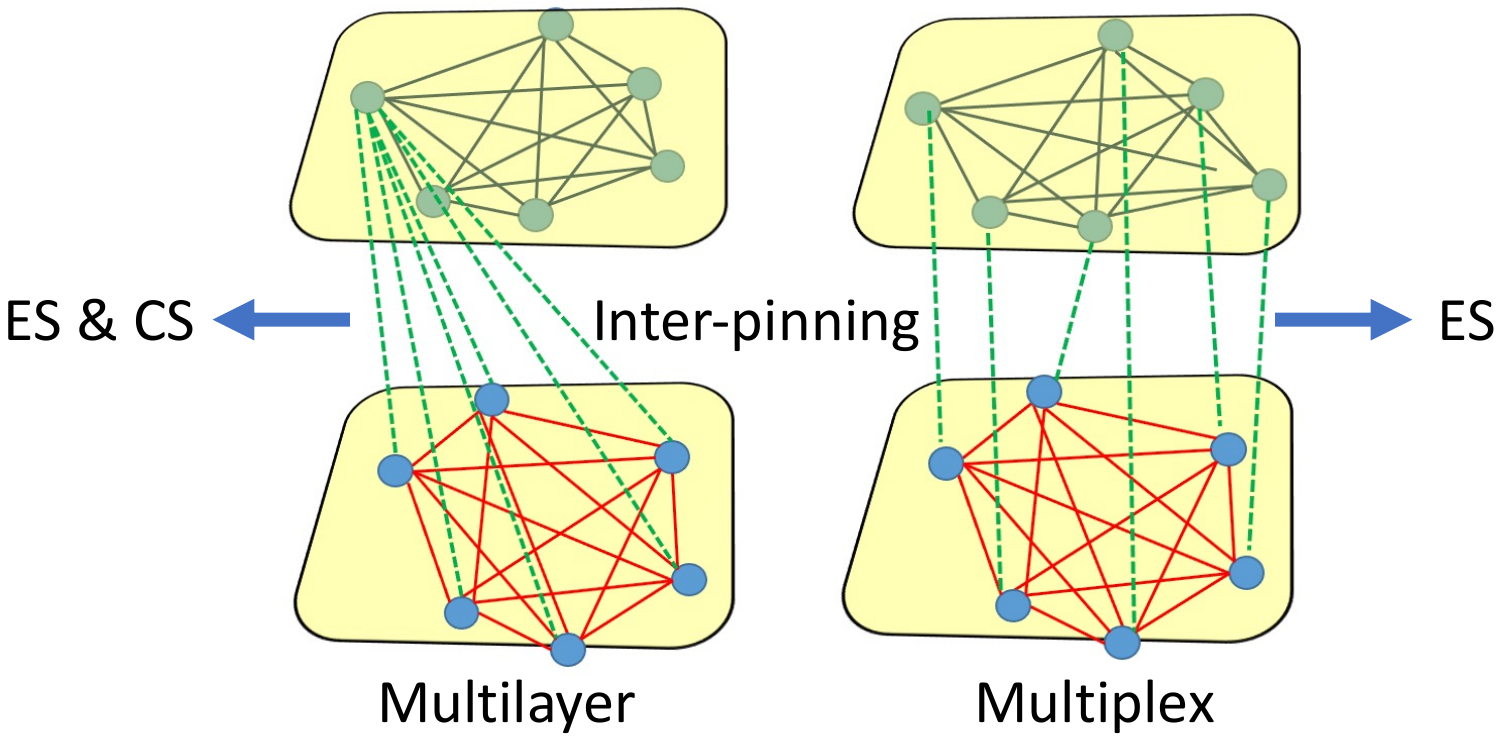}\\
	\caption{(Color online) Schematics of two all-to-all interconnected (green dashed lines) populations under the impression of interpinning, where ES and CS refer to explosive synchronization and chimera states, respectively. Multilayer formation: each node in a population is interconnected with all the nodes in the other population; for better clarity, only one node is shown connected with the other nodes in the other population. Multiplex formation: only mirror nodes are interconnected.}
	\label{figure1}
\end{figure}

Strogatz \textit{et al.}~\cite{strogatz1988,strogatz1989} showed that pinning the phases in networked oscillators to random phases leads to the emergence of the ES route. This work introduces a concept of multilayer interpinning (see Fig.~\ref{figure1}), which involves pairs of interconnecting nodes in two populations of nonidentical oscillators stuck at independent random phases. Such a scheme is more relevant when a particular impact of the interdependence, i.e., how activities of one network get affected by those of the other networks, is not known or decipherable from the available data, which is the issue for many complex systems (see Supplemental Material~\cite{SM}). Here we show that the interpinned multilayer network sports an interesting dynamical feature, i.e., the existence of chimera states (CS) during the explosive transition to synchronization and then desynchronization. The multilayer setup leads to the ES transitions in the two populations in succession, i.e., one population stays synchronous while the other stays asynchronous.\ We also covered multiplex interpinning (see Fig.~\ref{figure1}) in which the parallel nodes in the two populations are pinned to the same set of random phases. This setup induces ES transitions in the two populations concurrently, i.e., the occurrence of CS is not witnessed.\ Our investigation creates distinctions between the dynamical characteristics of the multilayer and multiplex interpinning.\\

\paragraph*{\bf{Dynamics on Multilayer Networks\label{sec:Model}}}
\noindent We begin with considering a multilayer network comprising two interacting nonidentical populations of the same number of nodes $N$. The evolution of phases $\theta_l^i$ ($i{=}1\dots N$) in either population $l\in\{1,2\}$ is governed by
\begin{equation}\label{eq:eqn}
	\dot\theta_l^i = \omega_l^i+\frac{\lambda}{N}\sum_{j=1}^N\sin(\theta_l^j-\theta_l^i)+ \frac{D}{N}\sum_{k=1}^N \sin(\theta_{l'}^k-\theta_l^i-\alpha^i),
\end{equation}
where the $\alpha^i$ are independent random phases uniformly distributed on the interval $\alpha^i\in[0,2\pi)$. The random pinning phases $\alpha^i$ corresponding to the mirror nodes $\{\theta_l^i,\theta_{l'}^i\}$ in the two populations are taken to be the same. Hence, the phase differences $(\theta^k_{l'}-\theta^i_l)$ of all the $k$ nodes in population $l'$ with an interconnected node $i$ in population $l$ are pinned at a random phase $\alpha^i$. For that matter, the third term tends to lead to a static disorderliness among all the interlayer phase differences. This static disorderliness creates frustration among the nodes in either population and hinders the synchronization process in them. Here, the interlayer coupling strength $D$ also serves as the pinning strength. The second term fosters intrapopulation coherence as the homogeneous coupling strength $\lambda=\lambda_l$ strengthens. The frequencies of the nodes in either population follow a uniform or symmetric distribution $g(\omega_l)$. Our aim is to comprehend how the behavior of phase synchronization in the two populations is influenced by the random interpinning. For that matter, the degree of synchronization in each population is determined by the order parameter defined as
\begin{equation}\label{eq:eqn1}
r_le^{i\psi_l} = \frac{1}{N}\sum_{j=1}^N e^{i\theta^j_l},
\end{equation}
where $\psi_l$ is the average phase of population $l$. A stationary value of $r(t)=r\simeq1$ implies coherence, whereas $r(t)\simeq0$ means complete incoherence.
\begin{figure}[t!]
	\centering
	\includegraphics[height=4cm,width=8cm]{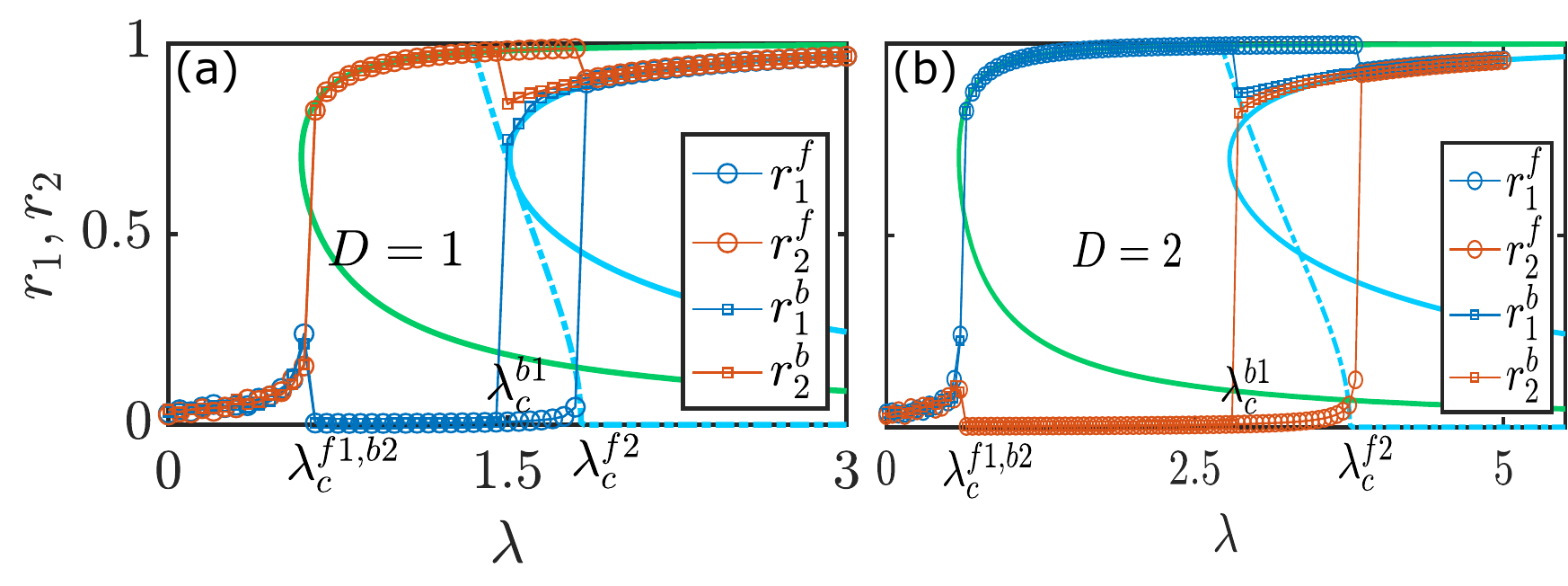}\\
	\includegraphics[height=2.5cm,width=8cm]{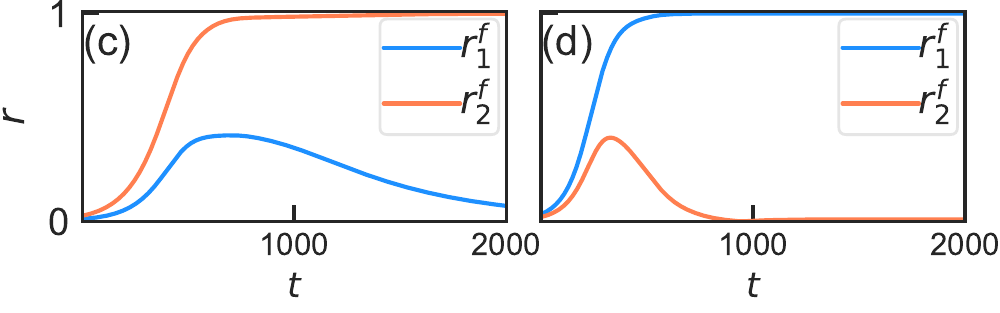}\\
	\includegraphics[height=3cm,width=8cm]{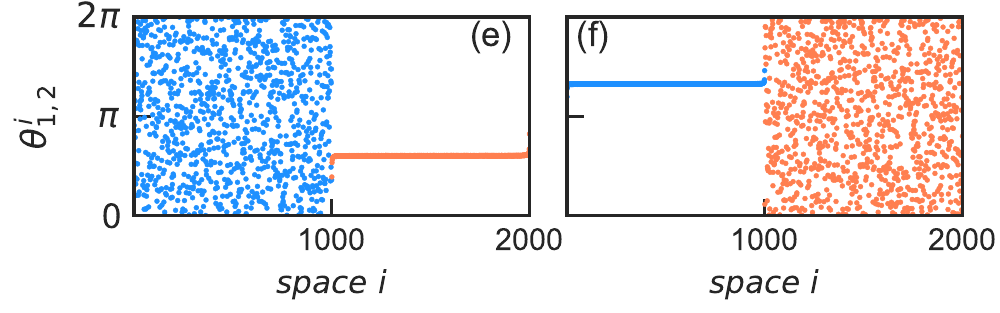}\\
	\vspace{-0.5cm}
	\caption{(Color online) (a), (b) $r_l{-}\lambda$ exhibiting ES transitions in a GC-GC multilayer network having uniform $\omega_l^{i}\in[-0.5,0.5]$. The green and cyan solid lines analytically match the stable synchronous solutions [Eq.~(\ref{eq:stable})] for both of the populations, while the cyan dashed line [Eq.~(\ref{eq:unstable})] elucidates the bistability in the solution for a population with hysteresis, as discussed later. Initial transients of forward $r_l^f$ for (c) $D{=}1$ and (d) $D{=}2$. Stationary phases of the two populations for (e) $D{=}1$ at $\lambda{=}1.6$ and (f) $D{=}2$ at $\lambda{=}2.5$. Here, $i{=}1\dots1000$ and $i{=}1001\dots2000$ belong to $\theta_1^i$ and $\theta_2^i$, respectively.}
	\label{figure2}
\end{figure}
\noindent We begin our investigation by constructing a multilayer network of two globally connected (GC) populations, each of size $N=1000$. The interlayer couplings between them are subject to random pinning, as discussed before. Different samples of natural frequencies for the two populations are selected from either uniform $g(\omega_l)\in[-\Delta,\Delta]$ or unimodal symmetric $g(\omega_l)$ with mean $0$. Distinct samples of phases for the two populations are drawn uniformly randomly on $[0, 2\pi)$. The phase dynamics of the multilayer network given by Eq.(\ref{eq:eqn}) is evolved using the Runge Kutta 4th order method with step size $\mathrm{dt}=0.01$. 
\paragraph{\textbf{Explosive synchronization (ES)}}
To witness the occurrence of ES, forward ($f$) and backward ($b$) phase transitions are observed by computing the order parameter against each adiabatically increasing or decreasing coupling strength $\lambda$ in the steps of $\mathrm{d}\lambda$, respectively.
In Fig.~\ref{figure2}, the order parameter corresponding to different values of the pinning strength $D$ is plotted for the forward and backward continuation in $\lambda$. It unveils that a sufficient pinning strength $D$ exerts frustration at the interconnected nodes and leads to a discontinuous transition in the two populations, accompanied by hysteresis. It is apparent that two sets of two distinct critical coupling strengths exist, one $\{\lambda_c^{f1},\lambda_c^{f2}\}$ for the forward abrupt transitions and the other $\{\lambda_c^{b1},\lambda_c^{b2}\}$ for the backward abrupt transitions as shown in Figs.~\ref{figure2}(a) and \ref{figure2}(b). At the first forward critical $\lambda_c^{f1}$, it is the initial condition dependence that one population experiences explosive transition while the other sees complete incoherence ($r_{l'}{\simeq}0$). The two populations remain in their respective states until second forward $\lambda_c^{f2}$ is reached. At $\lambda_c^{f2}$, the coherent population sees a marginal abrupt desynchronization while the incoherent population experiences explosive synchronization and traces the other synchronous population. In the backward transition, the subsequent abrupt desynchronization of the two synchronous populations takes place in a similar fashion. The population which synchronizes at $\lambda_c^{f2}$ is the first one to abruptly desynchronize at $\lambda_c^{b1}$, while the other population achieves a marginal abrupt gain in synchrony before desynchronizing at $\lambda_c^{b2}$. Next, the critical coupling points and the hysteresis width can be enhanced by increasing $D$ as it exerts even more frustration among the nodes, in turn entailing even larger values of $\lambda$ for the onset of an abrupt transition.

\paragraph{\textbf{Chimera states (CS)}}
 Here we emphasize the occurrence of the chimeric state in the two multilayered populations during the forward and backward phase transitions. In the multilayer formation, the two interlinked populations form the chimeric state, in which one population remains coherent while the other dwells in complete incoherence. In Figs.~\ref{figure2}(e) and \ref{figure2}(f), for instance, the stationary phases $\theta_l^i$ of the two populations are depicted exhibiting chimera states at $\lambda=2$ for different values of $D$. The region of chimeric occurrence spans from coupling strength $\lambda_{c}^{f1}$ to $\lambda_{c}^{f2}$ during the forward transition. The area sporting chimera states during the backward transition is stretched out in a relatively narrower region, beginning from coupling strength $\lambda_{c}^{b1}$ to $\lambda_c^{b2}$. For either population, whether it meets the coherence or incoherence in the chimera region depends purely on the sensitivity to the initial condition at critical point $\lambda_c^{f1}$ in the forward transition and $\lambda_c^{b1}$ in the backward transition. Figures~\ref{figure2}(c) and \ref{figure2}(d) exhibit the initial transients of forward $r_l$ for the two populations depending entirely upon the initial conditions at t=0. Further, the span of the existing chimera states during either the forward or backward transition gets augmented when the pinning strength is increased, as this, in turn, delays the values of $\lambda_c^{f2}$ and $\lambda_c^{b1}$ due to enhanced frustration among the nodes. The same is quite apparent from the $r_l-\lambda$ diagrams in Figs.~\ref{figure2}(a) and \ref{figure2}(b). The Supplemental Material~\cite{SM} reports (i) the occurrence of ES and CS for unimodal distribution $g(\omega_l)$ and (ii) the impact of randomly selected different fractions of the interpinned nodes on phase transition in the two populations.
\begin{figure}[t!]
	\centering
	\includegraphics[height=4cm,width=8.5cm]{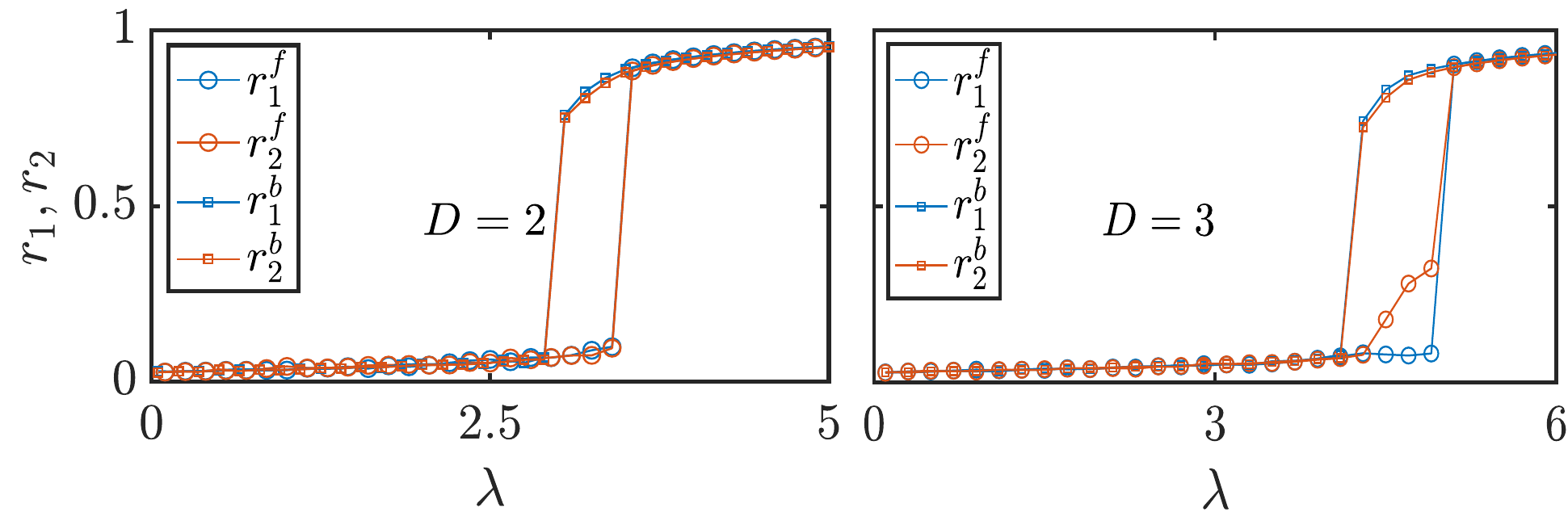}\\
	\vspace{-0.5cm}
	\caption{(Color online) $r_l$ for GC-GC multiplex network as a function of $\lambda$ simulated for uniform $g(\omega_l)$ with $\Delta{=}1$.}
	\label{figure4}
\end{figure}
\paragraph{\textbf{Theoretical predictions}}
In the thermodynamic limit $N\to\infty$, phases $\theta_l^i$ of the nodes in model (\ref{eq:eqn}) are continuous and a $2\pi$ periodic function such that $\alpha\to\theta_l^{\alpha}$~\cite{strogatz1989}. Each $\theta^i_l$ is associated with an $\alpha^i$; hence, relabeling of each $\theta_l^i$ with its corresponding $\alpha$ allows us to reexpress the model~(\ref{eq:eqn}) as \cite{strogatz1989}
\begin{multline}\label{eq:eqn2}
	\dot\theta_l^{\alpha} = \omega_l^{\alpha}+\frac{\lambda}{2\pi}\int_0^{2\pi}\mathrm{d\alpha'}\sin(\theta_l^{\alpha'}-\theta_l^{\alpha}) \\+ \frac{D}{2\pi}\int_0^{2\pi} \mathrm{d\alpha'}\sin(\theta_{l'}^{\alpha'}-\theta_l^{\alpha}-\alpha),
\end{multline}
where $l'\neq l;\ l,l'\in\{1,2\}$.
In the limit $N\to\infty$, the order parameter (\ref{eq:eqn1}) for a layer $l$ can be rewritten as \cite{strogatz1989}
\begin{align}\label{eq:eqn3}
r_le^{i\psi_l} = \frac{1}{2\pi}\int_{-\infty}^{\infty}\mathrm{d\omega_l}g(\omega_l)\int_0^{2\pi} e^{i\theta^{\alpha}_l} \mathrm{d\alpha}.
\end{align}
Model (\ref{eq:eqn2}) can be expressed in terms of mean-field parameters $r_l$ and $\psi_l$,
\begin{equation}\label{eq:eqn4}
	\dot\theta_l^{\alpha} = \omega_l^{\alpha}+\lambda r_l\sin(\psi_l-\theta_l^{\alpha})+ \mathrm{D} r_{l'}\sin(\psi_{l'}-\theta_l^{\alpha}-\alpha).
\end{equation}
Now considering $g(\omega_l)$ such that their mean frequencies $\Omega_l=0$, then $\psi_l=0$. The criteria for the synchronous states $\dot\theta^{\alpha}_l=0$ in either population then yields
\begin{eqnarray}\label{eq:eqn5}
e^{i\theta^{\alpha}_l} = \frac{i\omega^{\alpha}_l\pm\sqrt{|u+ve^{i\alpha}|^2-[\omega^{\alpha}_l]^2}} {u+ve^{i\alpha}},
\end{eqnarray}
where $u=\lambda r_l$ and $v = Dr_{l'}$. After substituting $e^{i\theta_l^{\alpha}}$ from Eq.(\ref{eq:eqn5}) into Eq.(\ref{eq:eqn3}), one obtains the following expression for the order parameter:
\begin{equation}\label{eq:eqn6}
	r_l = \frac{1}{2\pi}\int_{-\infty}^{\infty}\mathrm{d\omega_l} g(\omega_l)\int_0^{2\pi}\mathrm{d\alpha} \frac{[i\omega_l\pm\sqrt{|u+ve^{i\alpha}|^2-\omega_l^2}]} {u+ve^{i\alpha}}.
\end{equation}
We theoretically obtain the solutions for synchronous states by considering uniform $g(\omega_l)=\frac{1}{2\gamma}\ \mbox{for}\ w_l^{\alpha}\in[-\gamma, \gamma]$ such that $\Omega_l=0$. For the uniform $g(\omega_l)$, the first part of the integration in the order parameter~(\ref{eq:eqn6}) vanishes,
\begin{eqnarray}
	\frac{1}{2\pi}\int_{-\gamma}^{\gamma}\frac{\mathrm{d\omega_l}}{2\gamma}\int_0^{2\pi}\mathrm{d\alpha} \frac{i\omega_l}{u+ve^{i\alpha}} = 0,
\end{eqnarray}
and only the second part accounts for the order parameter. Since we must have $r_l>0$, the $+$ sign in the second term in Eq.~(\ref{eq:eqn6}) is taken into account and then $r_l$ can be reexpressed in terms of $z=u/v$ as
\begin{align}
	r_l = \frac{1}{2\pi}\int_{-\gamma}^{\gamma}\frac{\mathrm{d\omega_l}}{2\gamma} \int_0^{2\pi}\mathrm{d\alpha} \frac{\sqrt{|z+e^{i\alpha}|^2-[\omega_l/v]^2}} {z+e^{i\alpha}}.
\end{align}
After carrying out some mathematical simplifications, the real part of the order parameter is expressed as
\begin{multline}\label{eq:stable}
	r_l = \frac{1}{2\pi}\int_{-\gamma}^{\gamma}\frac{\mathrm{d\omega_l}}{2\gamma} \int_0^{2\pi}\mathrm{d\alpha} \frac{\sqrt{z^2+2z\cos\alpha+1-[\omega_l/v]^2}} {z^2+2z\cos\alpha+1}\\ (z+\cos\alpha).
\end{multline}
Here, $r_l=0$ is one of the solutions of Eq.~(\ref{eq:stable}) for $z=0$, i.e., $u=0$ and $v\neq0$. The bifurcating solutions in the vicinity of $z=0$ $(z\to0)$ are obtained by the series expansion of Eq.~(\ref{eq:stable}) for $0<z<1$:
\begin{eqnarray}\label{eq:unstable}
	r_l = \frac{z^3}{16[1-\frac{\gamma^2}{v^2}]^{3/2}}+\frac{u}{2\gamma}\arcsin\bigg[\frac{\gamma}{v}\bigg]+\mathcal{O}(z^4).
\end{eqnarray}
\begin{figure}[t!]
	\centering
	\begin{tabular}{cc}
	%\hspace{-0.5cm}
	\includegraphics[height=4cm,width=4cm]{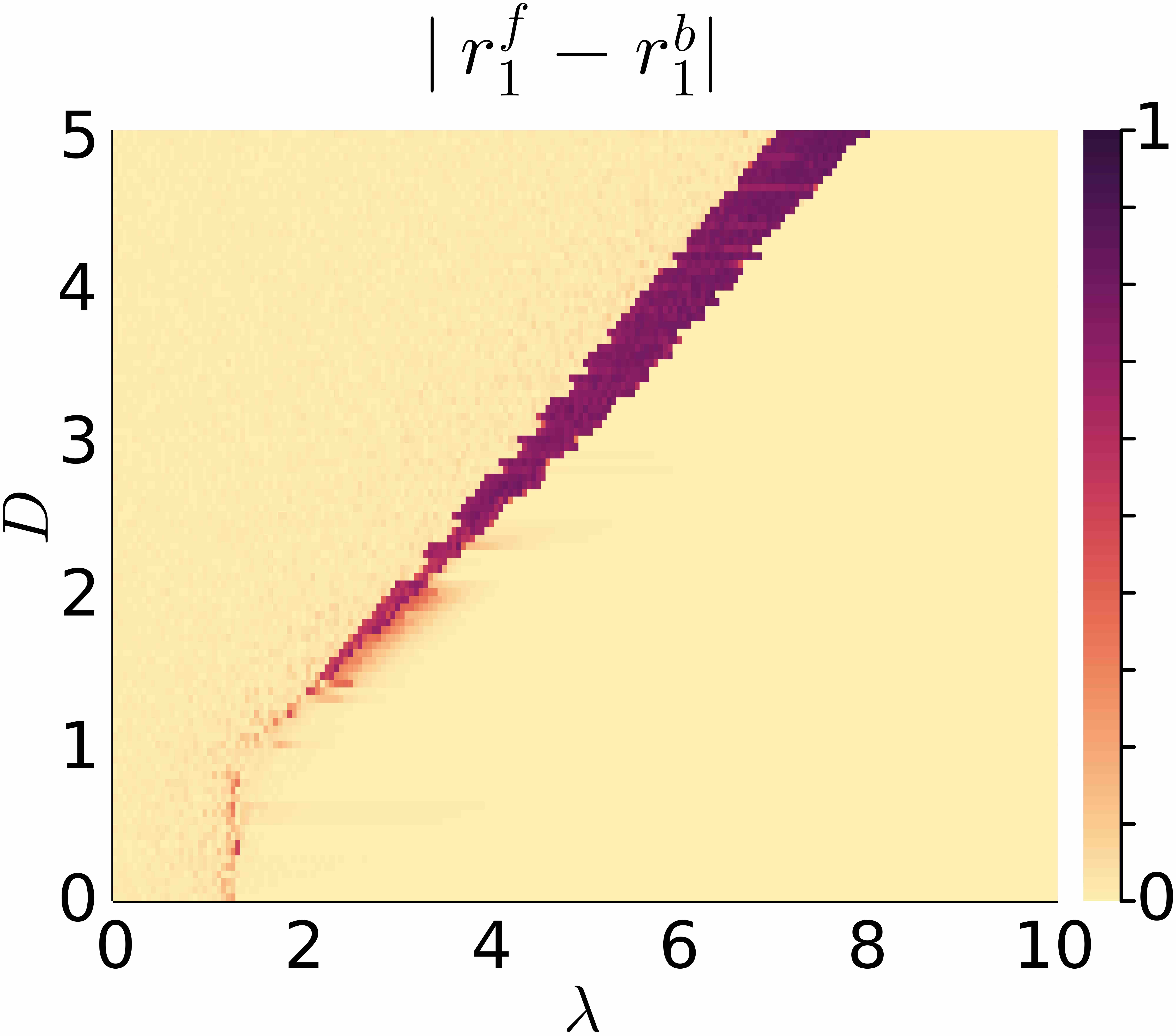}&
	\includegraphics[height=4cm,width=4cm]{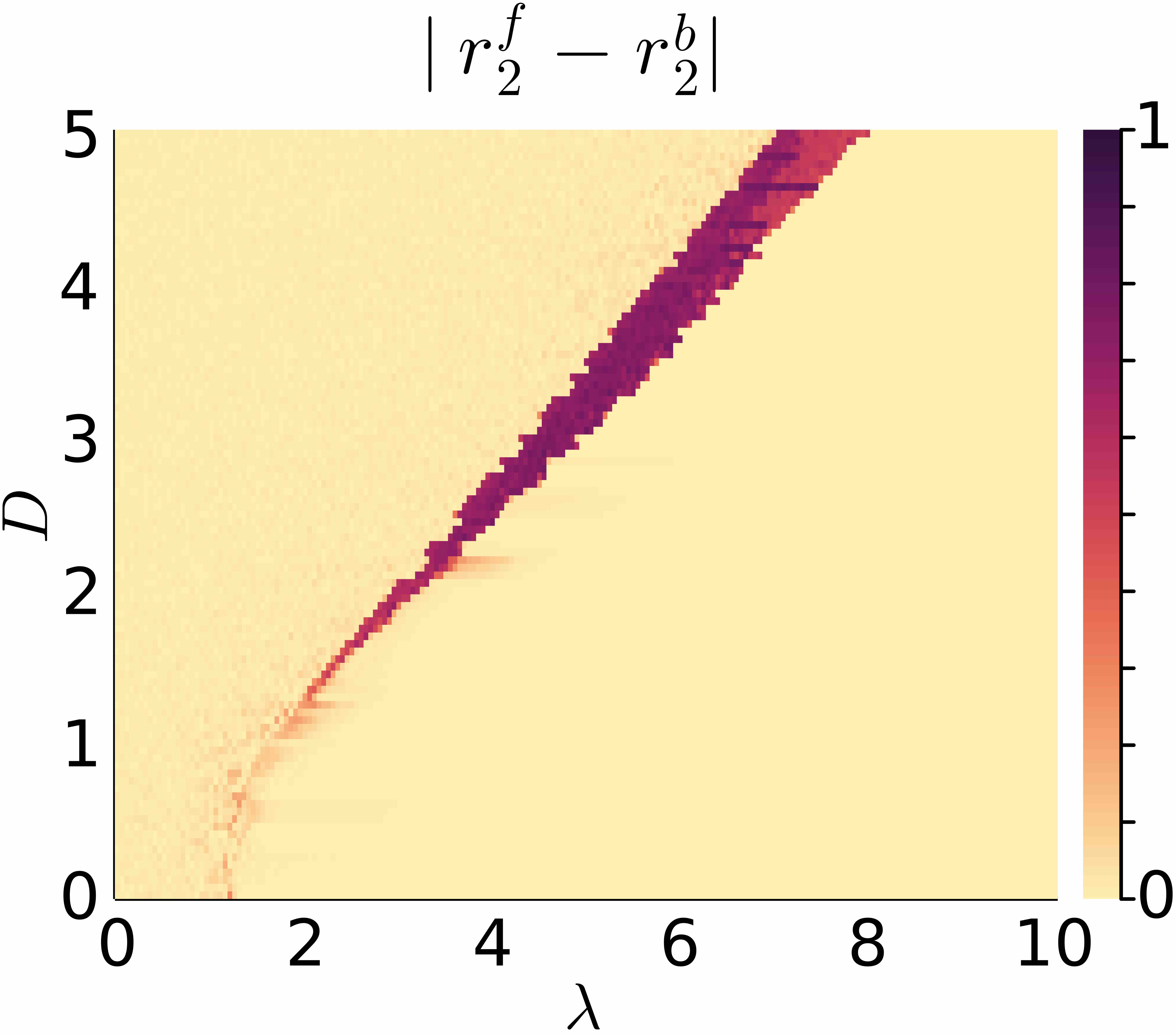}\\
	\end{tabular}{}
	\vspace{-0.5cm}
	\caption{(Color online) Abrupt jump size $|r_l^f-r_l^b|$; $l\in[1,2]$ and hysteresis width in $D-\lambda$ space for GC-GC multiplex network ($N=1000$) having uniform frequencies with $\Delta=1$.}
	\label{figure5}
\end{figure}
The solutions for the order parameter given by Eqs.~(\ref{eq:stable}) and (\ref{eq:unstable}) are depicted, respectively, by cyan and green solid lines, and cyan dashed lines in Figs.~\ref{figure2}(a) and \ref{figure2}(b) for different values of $D$. In Fig.~\ref{figure2}(a) for $D{=}1$, Eq.~(\ref{eq:stable}) yields stable coherent traces for $r_2^b$ after supplying $v{=}r_1^b{\simeq}0.004$ (from simulation) at $\lambda_c^{b2}$, while Eq.~(\ref{eq:unstable}) does not hold any feasible solution as no hysteresis exists for $r_2^{f,b}$. Hence, at $\lambda_c^{b2}$, $v{=}r_1^b{\simeq}0$ yields a large $u{=}r_2^b$, which defines one boundary for chimeric states.
Further, the stable and unstable traces for $r_1^b$ using Eq.~(\ref{eq:stable}) and Eq.~(\ref{eq:unstable}), respectively, are obtained after supplying $v{=}r_2^b{\simeq}0.978$ (from simulation) at $\lambda_c^{b1}$.\ Hence, at $\lambda_c^{b1}$, $v{=}r_2^b{\simeq}1$ yields $u{=}r_1^b{\simeq}0$, which defines the other boundary for chimeric states.
During the forward transition, $r_1^f{=}0$ [from Eq.~(\ref{eq:stable})], the stable fixed point solution for $\lambda{<}\lambda_c^{f2}$ becomes unstable at $\lambda_c^{f2}$ when it coalesces with the unstable fixed point given by Eq.~(\ref{eq:unstable}). Hence, for $\lambda{>}\lambda_c^{f2}$, the incoherence is lost (unstable $r_1^f{=}0$), as shown by the cyan dashed line Eq.~(\ref{eq:unstable}), and a stable solution at large $r_1^f$ abruptly emerges.\ During the backward transition, the stable fixed point $r_1^b$ [Eq.~(\ref{eq:stable}] and the unstable fixed point [Eq.~(\ref{eq:unstable})] coalesce at $\lambda_c^{b1}$, and both are then annihilated; in turn, the coherent trace $r_1^b$ is lost for $\lambda{<}\lambda_c^{b1}$. Only the stable incoherent solution $r_1^b\simeq0$ then exists for $\lambda{<}\lambda_c^{b1}$.\\
In similar fashion for $D{=}2$ [see Fig.~\ref{figure2}(b)], the stable coherent and unstable traces for $r_1^b$ and $r_2^b$ are obtained using Eqs.~(\ref{eq:stable}) and (\ref{eq:unstable}) after supplying $v{=}r_2^b{\simeq}0.002$ (at $\lambda_c^{b2}$) and $v{=}r_1^b{\simeq}0.99$ (at $\lambda_c^{b1}$), respectively.\\
Thus, the theoretical predictions given by Eqs.~(\ref{eq:stable}) and (\ref{eq:unstable}) successfully elucidate the subsequent onset of explosive synchronization and desynchronization transitions in the two populations with defined boundaries for chimera states as in the critical coupling strengths.\\

\paragraph*{\bf{Dynamics on Multiplex Networks}}
Next, we treat a multiplex framework of the model given in Eq.~(\ref{eq:eqn}), which considers interactions only between the mirror adjacent nodes in the two populations. The evolution of phases in the multiplex network possessing random interpinning in the mirror nodes is expressed as
\begin{equation}\label{eq:eqn_}
	\dot\theta_l^i = \omega_l^i+\frac{\lambda}{N}\sum_{j=1}^N\sin(\theta_l^j-\theta_l^i)+ D\sin(\theta_{l'}^i-\theta_l^i-\alpha^i).
\end{equation}
Synchronization diagrams for such multiplex network consisting of two GC populations are shown in Fig.~\ref{figure4} for different values of $D$. Both the forward critical coupling strength and the hysteresis width increase with the increase in $D$. Also, no initial condition dependence of the order parameter is witnessed for the multiplexed populations, unlike what we witnessed in the case of the multilayered populations.

\paragraph{\textbf{Phase plot in $D-\lambda$ space}} To have a complete picture of the nature of the transition with change in the pinning strength, we draw a phase plot in the $D-\lambda$ space for each layer.
The color profile in the $D-\lambda$ space in Fig.~\ref{figure5} illustrates the abrupt jump size $|r_l^f-r_l^b|;\ l\in[1,2]$ for a GC-GC multiplexes having natural frequencies drawn from a uniform distribution. A profound distinction in color between the hysteresis region and asynchronous or synchronous region can be witnessed for pinning strength $D>1$. The magnitude of forward and backward critical coupling strength and hysteresis width corresponding to a pinning strength $D$ can also be extracted from the $D-\lambda$ phase plots.
\begin{figure}[t!]
	\centering
	\begin{tabular}{cc}
	\hspace{-0.5cm}
	\includegraphics[height=3.5cm,width=8.5cm]{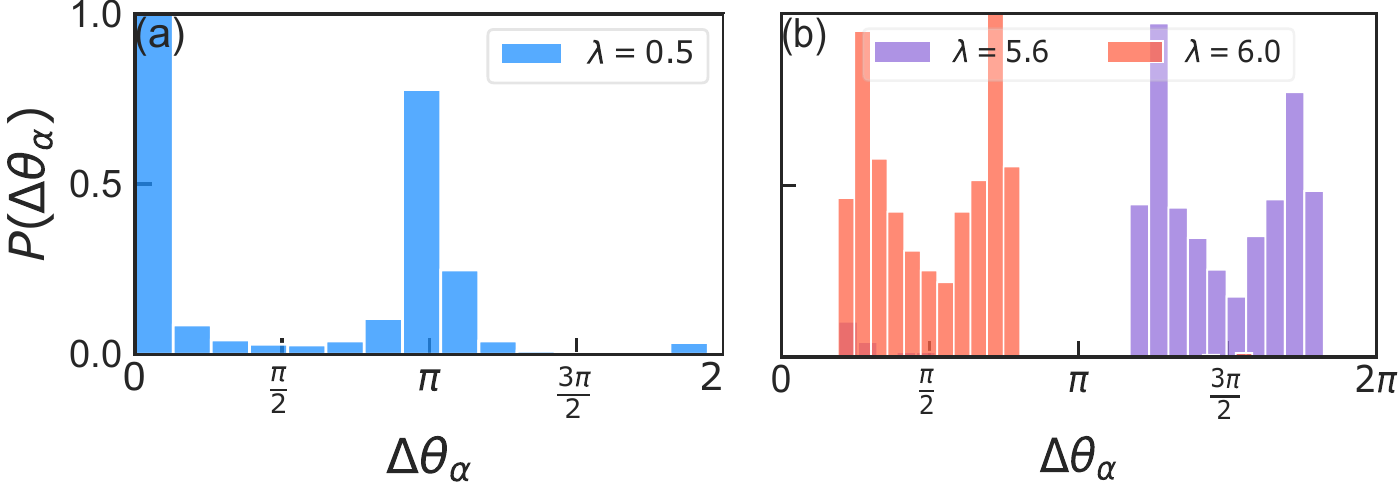}\\
	\includegraphics[height=3.5cm,width=8.5cm]{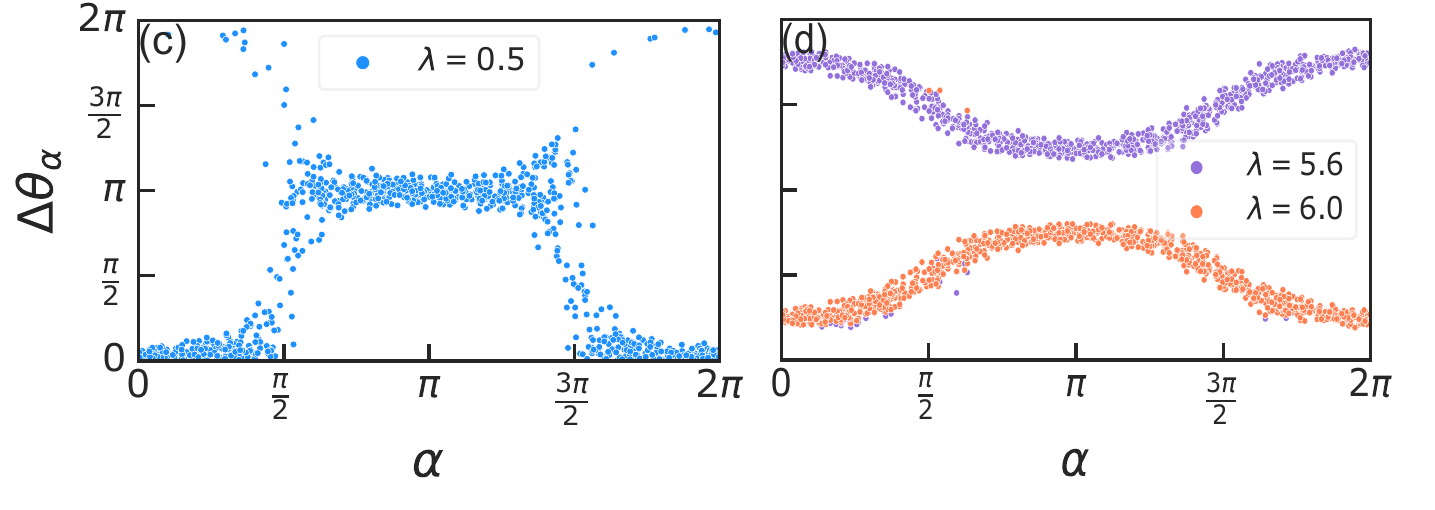}\\
	\end{tabular}{}
	\vspace{-0.5cm}
	\caption{(Color online) Phase distribution of $\Delta\theta^{\alpha}{=}|\theta_1^{\alpha}-\theta_2^{\alpha}|$, and $\Delta\theta^{\alpha}$ as a function of $\alpha$ for different values of $\lambda$ for the GC-GC multiplex network with $D{=}3$ and uniform $\omega_l^{\alpha}$ with $\Delta{=}1$.}
	\label{figure6}
\end{figure}

\paragraph{\textbf{Distribution of difference of mirror phases}}
\noindent Here, we investigate the microscopic dynamics of $\Delta\theta^{\alpha}{=}|\theta_1^{\alpha}-\theta_2^{\alpha}|$, the phase difference between the mirror nodes.\ Figure~\ref{figure6} illustrates the distribution $P(\Delta\theta^{\alpha})$ in the asynchronous and synchronous states which belong to the $r-\lambda$ profile for $D{=}3$ in Fig.~\ref{figure4}. 
The distribution $P(\Delta\theta^{\alpha})$ for any $\lambda<\lambda_c^f$ exhibits two peaks at $\Delta\theta^{\alpha}{=}0$ and $\Delta\theta^{\alpha}{=}\pi$, as shown in Fig.~\ref{figure6}(a).\ It implies that for any $\lambda$ belonging to the asynchronous state, the $N$ sized population of $\Delta\theta^{\alpha}$ is segregated notably in two clusters, one at $0$ or $2\pi$, and the other at $\pi$, with a few sparsely populated elsewhere. On the other hand, in the synchronous state $\lambda>\lambda_c^f$, the $P(\Delta\theta^{\alpha})$ exhibits bimodal peaks with their minima located at either $\pi/2$ or $3\pi/2$ [see Fig.~\ref{figure6}(b)], and hence the two peaks are located at a spread of $\pi/4$ on either side of the minima.
Furthermore, we study $\Delta\theta^{\alpha}$ as a function of $\alpha$ in Fig.~\ref{figure6} for different values of $\lambda$. For any $\lambda<\lambda_c^f$ [see Fig.~\ref{figure6}(c)], the nodes whose initial independent pinning phases are bounded within $\alpha\in[\pi/2,3\pi/2]$ achieve a steady state around $\Delta\theta^{\alpha}{=}\pi$, whereas the nodes whose $\alpha{\in}[0,\pi/2]$ or $\alpha{\in}[3\pi/2,2\pi]$ settle on about $\Delta\theta^{\alpha}{=}0,2\pi$ in the steady state. Nevertheless, for any $\lambda>\lambda_c^f$ [see Fig.~\ref{figure6}(d)], the steady state values of $\Delta\theta^{\alpha}$ are spread between either $[5\pi/4$, $7\pi/4]$ or [$\pi/4$, $3\pi/4$] depending on the value of $\lambda$. The stationary population of $\Delta\theta^{\alpha}$ in the asynchronous and synchronous state corroborates with the findings for $P(\Delta\theta^{\alpha})$.\\

\begin{figure}[t!]
	\centering
	\begin{tabular}{cc}
	\includegraphics[height=3.8cm,width=8.1cm]{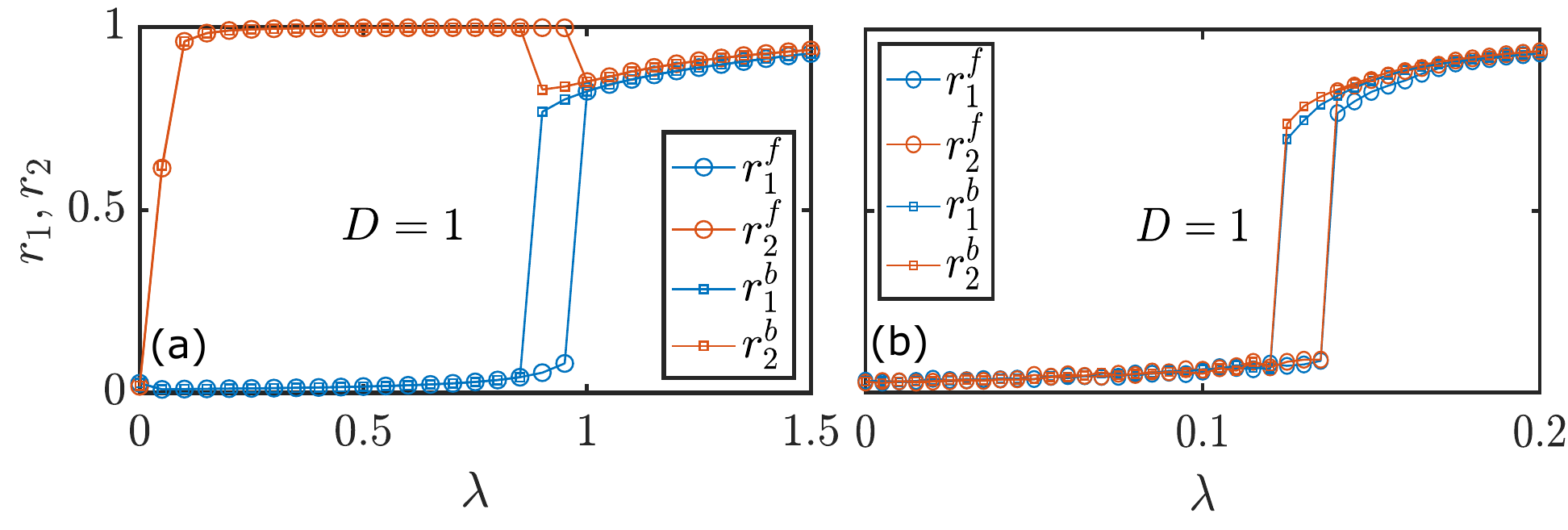}\\[-2ex]
	\includegraphics[height=4cm,width=8.2cm]{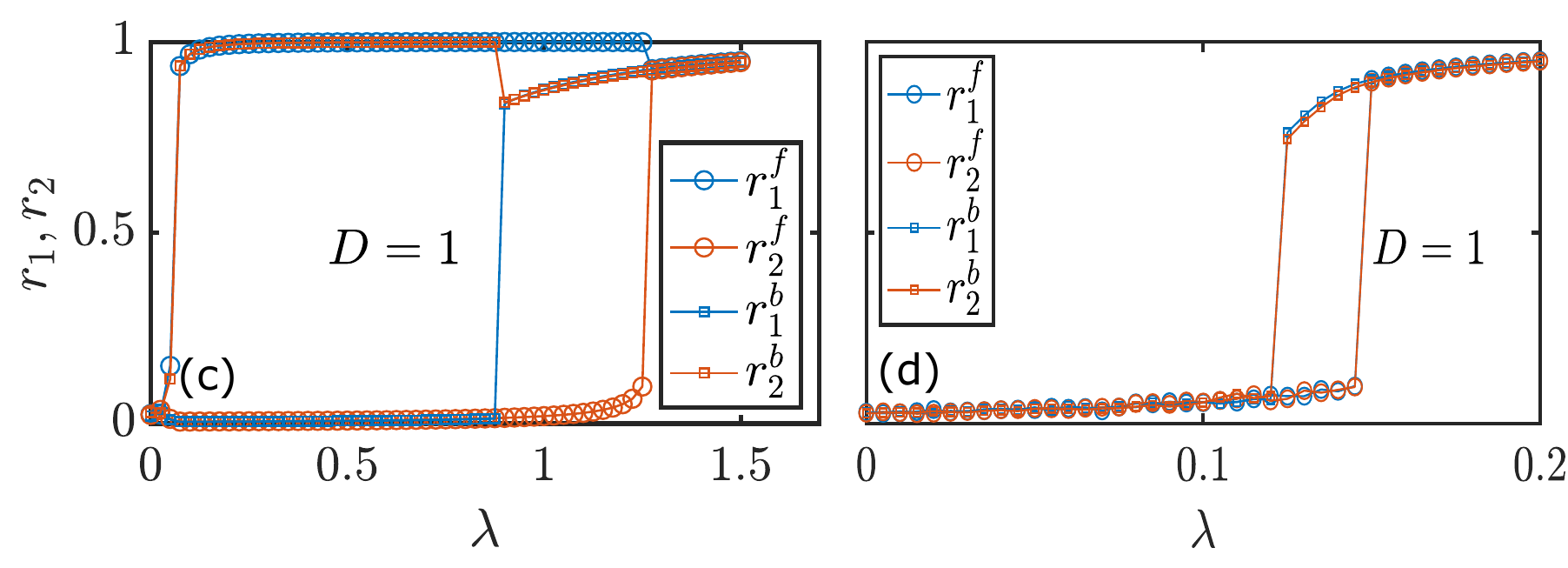}\\
	\end{tabular}
	\vspace{-0.5cm}
	\caption{(Color online) $r_l{-}\lambda$ profiles for ER-ER and WS-WS populations $(\langle k_1\rangle{=}\langle k_2\rangle{=}12; N{=}1000$, $\omega_l^i{\in}[-0.5,0.5])$.\ For multilayer networks with (a) ER and (c) WS interconnectivity $\langle k_I\rangle{=}8$. For multiplex networks, (b) ER-ER and (d) WS-WS.}
	\label{figure7}
\end{figure}	
\paragraph*{\bf{The robustness of interpinning prescription against populations' topology}}
The interpinning prescription to the multilayer and multiplex networks also successfully applies to the populations' connectivity manifesting a topology other than GC-GC. We demonstrate this representing the two populations by ER-ER (Erd\"os R\'enyi) random~\cite{Erdos1960} and WS-WS (Watts-Strogatz) small-world~\cite{Watts} networks interconnected in multilayer formation,
	\begin{equation}\label{eq:SM_multi}
	\dot\theta_l^i = \omega_l^i+\lambda\sum_{j=1}^N A_l^{ij}\sin(\theta_l^j-\theta_l^i)+ D\sum_{k=1}^N A_I^{ik}\sin(\theta_{l'}^k-\theta_l^i-\alpha^i),
	\end{equation}
	and multiplex formation,
	\begin{equation}\label{eq:SM_mplex}
	\dot\theta_l^i = \omega_l^i+\lambda\sum_{j=1}^N A_l^{ij}\sin(\theta_l^j-\theta_l^i)+ D\sin(\theta_{l'}^i-\theta_l^i-\alpha^i),
	\end{equation}
	where $A_l; l{\in}\{1,2\}$ represents the adjacency matrix of a population.\ The multiplex formation given by Eq.~(\ref{eq:SM_mplex}) has the interconnections only with the mirror nodes, while the interconnections in the multilayer formation given by Eq.~(\ref{eq:SM_multi}) manifest ER (WS) network connectivity $A_I$ with average degree $\langle k_I\rangle$.\ The ER-ER (WS-WS) populations interpinned in multilayer and multiplex formations also exhibit ES transitions successively and simultaneously, as shown in Fig.~\ref{figure7}(a) [Fig.~\ref{figure7}(c)] and Fig.~\ref{figure7}(b) [Fig.~\ref{figure7}(d)], respectively.\ Nonetheless, the multiplex and multilayer formations of BA-BA populations, where BA (Barab\'asi-Albert) denotes scale-free topology~\cite{barabasi}, do not exhibit an ES transition because of the heterogeneity of BA topology.

\paragraph*{\bf{Conclusion}} In summary, we considered multilayer networks in which two populations are randomly interpinned. Such an arrangement leads to the explosive synchronization in the two populations in succession. After the initiation of an explosive transition in one population, the multilayer networks stay in the coexisting state of coherent and incoherent populations until the other population also undergoes the explosive transition. Such chimeric pattern in the two populations is witnessed during their explosive transition to synchronization and then desynchronization. Also, the abrupt synchronization and desynchronization transitions are found to be initial condition dependent. The analytical predictions for the order parameter are also provided, which fall into good agreement with the numerical estimations. We also explored the phase transition in the multiplex network in which only mirror nodes in the two populations are randomly interpinned. In the multiplex network, both of the populations espoused explosive transitions route to synchronization simultaneously. The order parameters of the populations in multiplex formation did not exhibit any chimeric state and initial condition dependence.
One can find an analogy of the abrupt onset (explosive transition) to the chimeric state and then abrupt return to normalcy in the multilayer formation of two populations with the sudden onset and offset of the focal seizures in the brain in which only a part of the brain experiences a seizure episode, while the other part functions normally~\cite{Loddenkemper2005}.

\begin{acknowledgments}
S.J. is thankful for the financial support of the Government of India, Council of Scientific \& Industrial Research (CSIR) Grant No.\ 25(0293)/18/EMR-II and Board of Research in Nuclear Sciences (BRNS) Grant No.\ 37(3)/14/11/2018-BRNS/37131.\ A.D.K. acknowledges the Government of India, Council of Scientific \& Industrial Research (CSIR) Grant No.\ 25(0293)/18/EMR-II for the RA fellowship.
\end{acknowledgments}

%\bibliographystyle{apsrev4-2}
%\bibliography{ipinning}

\end{document}

\title{Supplemental Material: Explosive Synchronization and Chimera in Inter-pinned Multilayer Networks}

\author{Ajay Deep Kachhvah$^1$, and Sarika Jalan$^{1,2}$}
\affiliation{1. Complex Systems Lab, Department of Physics, Indian Institute of Technology Indore, Khandwa Road, Simrol, Indore-453552, India}
\affiliation{2. Department of Biosciences and Biomedical Engineering, Indian Institute of Technology Indore, Khandwa Road, Simrol, Indore-453552, India}

\maketitle

\subsection{\textbf{Explosive synchronization and chimera state for unimodal frequency distributions}}
\begin{figure}[ht]
	\centering
	\begin{tabular}{cc}
	\includegraphics[height=4.1cm,width=8cm]{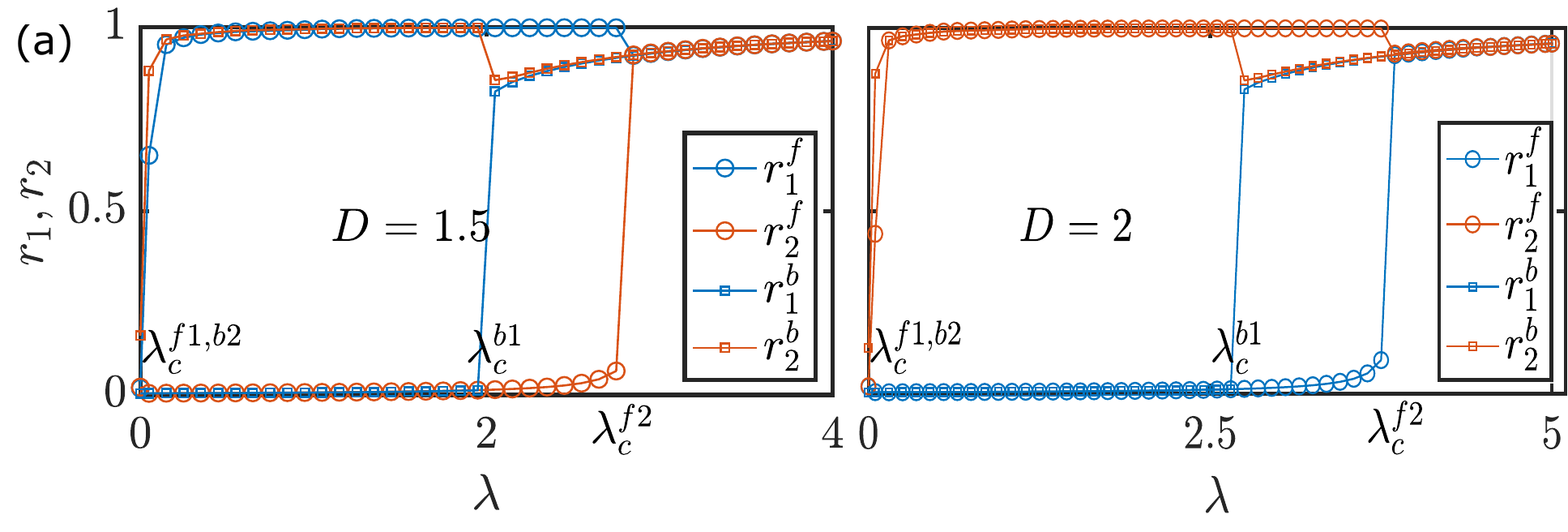}&
	\includegraphics[height=4cm,width=8cm]{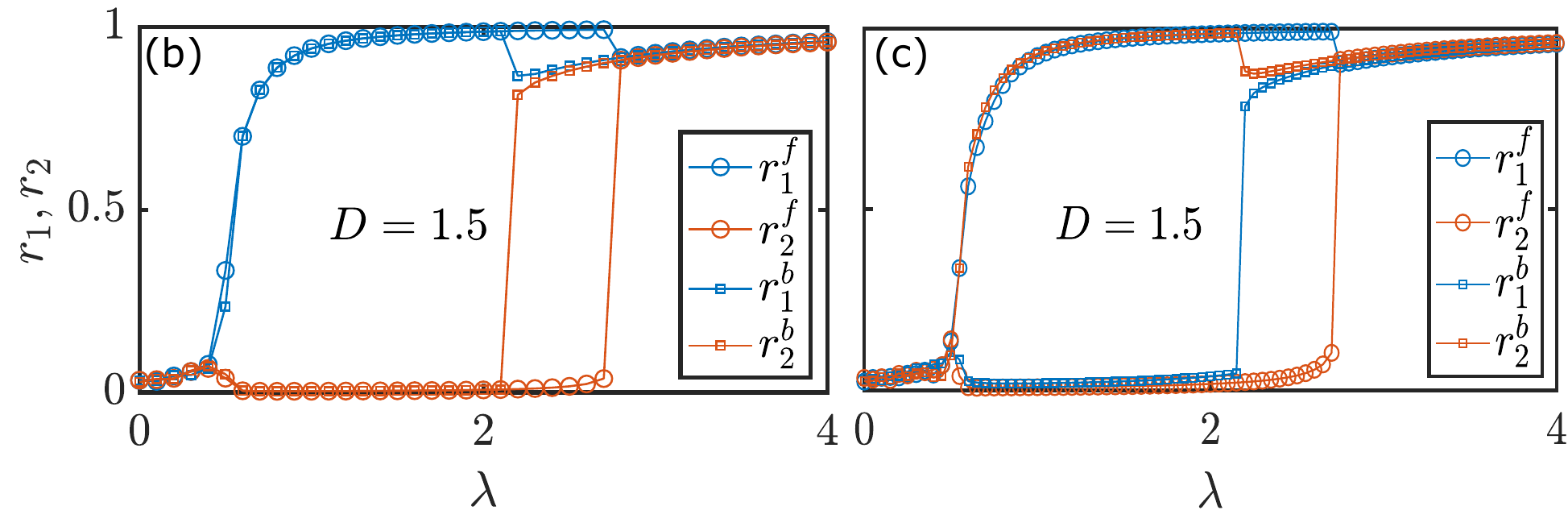}\\
	\includegraphics[height=4cm,width=8cm]{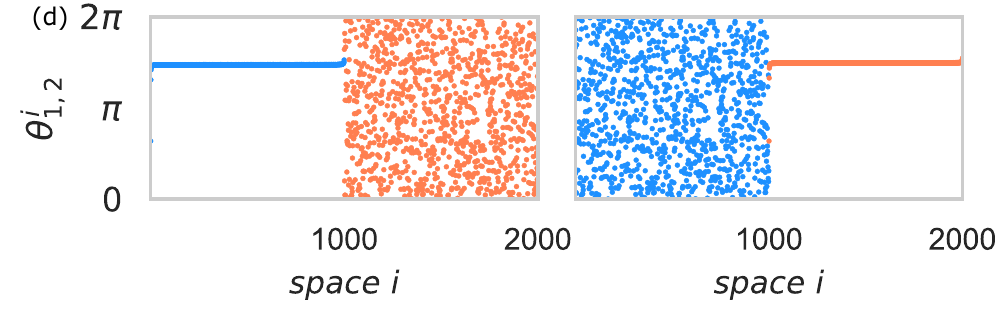}&
	\includegraphics[height=4cm,width=8cm]{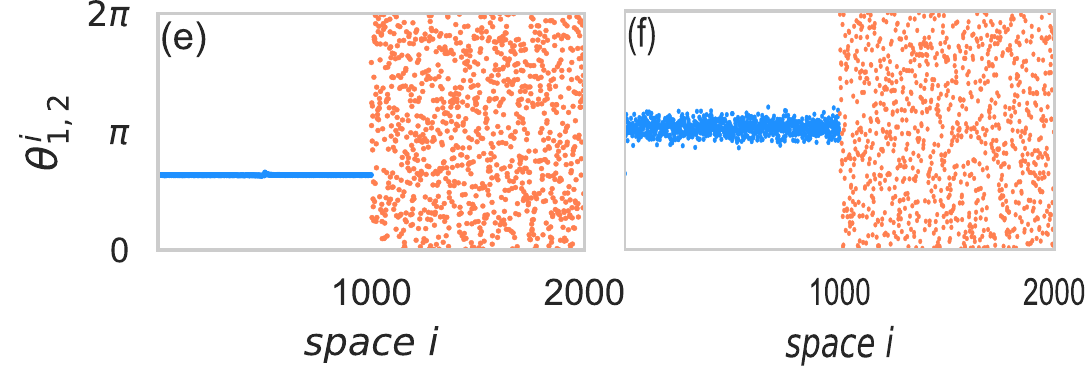}\\
	\end{tabular}{}
	\caption{(Color online) $r_l-\lambda$ profiles of GC-GC multilayer networks having (a) Lorentzian $g(\omega_l)$ with $\omega_1^i{=}\omega_2^i$, and Gaussian $g(\omega_l)$ with (b) $\omega_1^i{=}\omega_2^i$ and (c)  $\omega_1^i{\neq}\omega_2^i$. Stationary phases $\theta_1^i$ ($i{=}1\dots1000$) and $\theta_2^i$ ($i{=}1001\dots2000$) at $\lambda=2$ corresponding to (d) Lorentzian $g(\omega_l)$ for $D{=}1.5$ (left plot) and $D{=}2$ (right plot), and Gaussian $g(\omega_l)$ for (e) $\omega_1^i{=}\omega_2^i$ and (f) $\omega_1^i{\neq}\omega_2^i$.}
	\label{SM_figure1}
\end{figure}

We have already discussed the interesting case of initial condition dependence of the transition route for $r_l$ at the forward critical point $\lambda_{c}^{f1}$ in the main. The sensitivity to initial condition of $r_l$ during backward transition is demonstrated in Fig.~\ref{SM_figure1}(a) for $\omega_1^i{=}\omega_2^i$ following Lorentzian distribution $g(\omega_l)=\frac{\Delta}{\pi(\omega_l^2+\Delta^2)}$, where the half width at half maximum $\Delta{=}1$. It is shown for $D=2$ that $r_1^b$ forms the hysteresis loop with $r_1^f$, and $r_2^b$ also traces back $r_2^f$ branch from the abrupt transition point $\lambda_{c}^{b1}$. However as shown for $D=1.5$, $r_2^b$ does not form the hysteresis loop with $r_2^f$ instead $r_1^b$ does, and $r_2^b$ instead traces the $r_1^f$ branch from the transition point $\lambda_{c}^{b1}$. The reason for such random behavior of $r_l^b$ in adopting either of the backward transition routes is its sensitivity to initial condition at $\lambda_c^{b1}$. The initial condition dependence of the forward and backward $r_l$ has been witnessed for different values of $D$ and different types of distribution. Here we also depict the stationary phases $\theta_l^i$ of the two populations forming chimera states at $\lambda=2$ for different values of $D$ (see Fig.~\ref{SM_figure1}(d)). The subsequent occurrence of ES in $r_l-\lambda$ transition and chimera states are also witnessed for Gaussian distribution $g(\omega_l){=}\frac{1}{\Delta\sqrt{2\pi}}\exp(-\omega_l^2/2\Delta^2); \Delta{=}1$ for $\omega_1^i{=}\omega_2^i$ (see Fig.~\ref{SM_figure1}(b,e)). Chimera like state with one population in partial synchrony is also witnessed for $\omega_1^i{\neq}\omega_2^i$ as shown in Fig.~\ref{SM_figure1}(f) corresponding to $r_l-\lambda$ profile (see Fig.~\ref{SM_figure1}(c)). Therefore, the proposed model is also valid for symmetric unimodal $g(\omega_l)$ in giving rise to ES and CS.

\subsection{Impact of the fraction $f$ of inter-pinned nodes}
\begin{figure}[ht]
	\centering
	\begin{tabular}{cccc}
	\includegraphics[height=4cm,width=4cm]{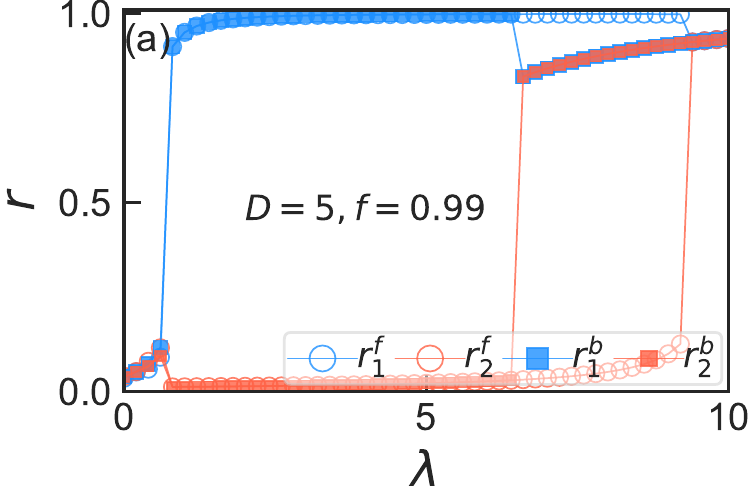}&
	\includegraphics[height=4cm,width=4cm]{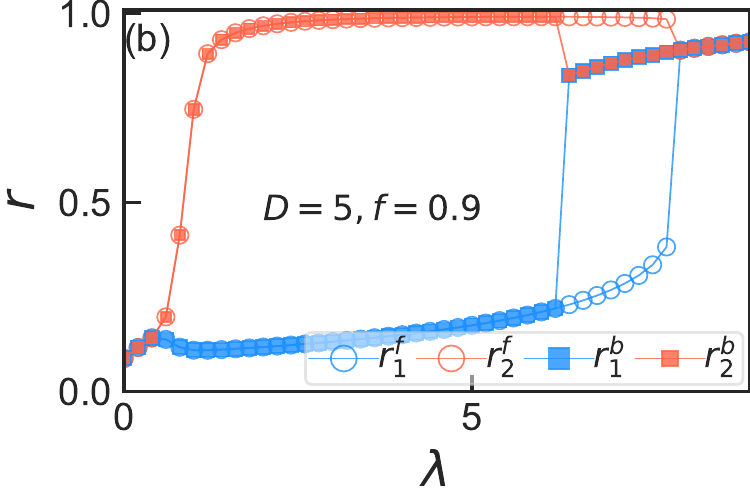}&
	\includegraphics[height=4cm,width=4cm]{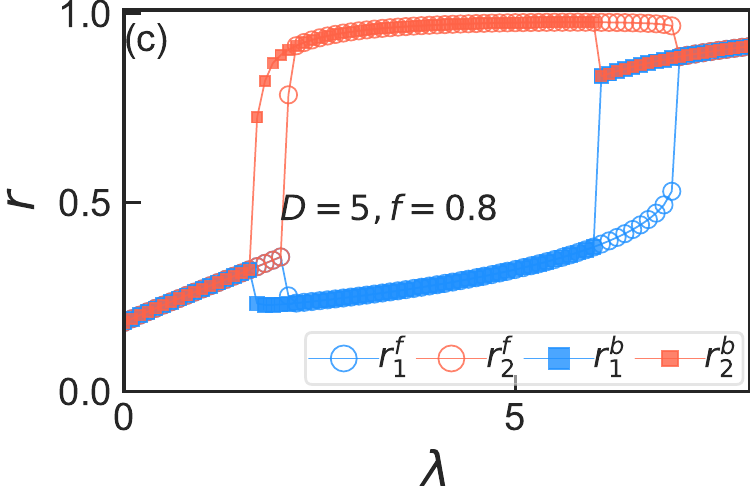}&
	\includegraphics[height=4cm,width=4cm]{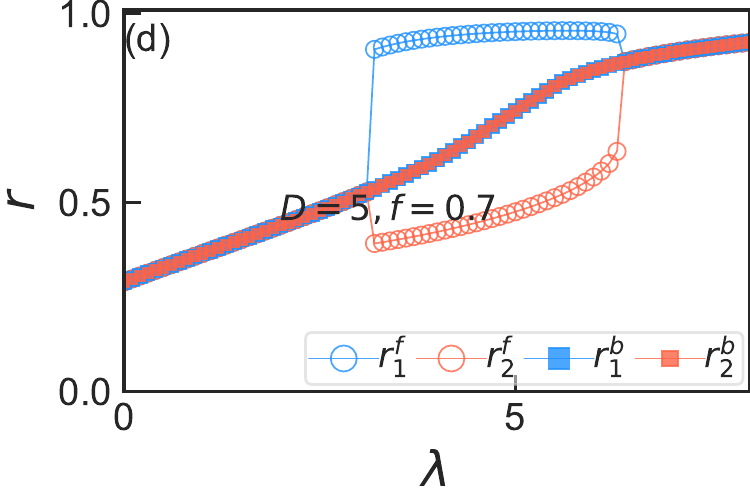}\\
	\end{tabular}{}
	\caption{(Color online) $r_l-\lambda$ profiles for GC-GC multilayer networks (for $D{=}5$) composed of different fraction values $f$ of the pinned nodes.}
	\label{SM_figure2}
\end{figure}

Here we discuss the case of a GC-GC multilayer network in which a fraction $f$ of the randomly selected nodes in both the populations are inter-pinned. The evolution of phases in such a multilayer formation is governed by
\begin{equation}\label{eq:eqn}
	\dot\theta_l^i = \omega_l^i+\frac{\lambda}{N}\sum_{j=1}^N\sin(\theta_l^j-\theta_l^i)+ f \frac{D}{N}\sum_{k=1}^N \sin(\theta_{l'}^k-\theta_l^i-\alpha^i) + (1-f) \frac{D}{N}\sum_{k=1}^N \sin(\theta_{l'}^k-\theta_l^i).
\end{equation}
In Fig.~\ref{SM_figure2}, synchronization profiles for different values of the fraction $f$ of the inter-pinned nodes are shown for sufficiently strong pinning strength $D=5$. As the fraction $f$ of the inter-pinned nodes is decreased, the hysteresis width for the population that synchronizes later gradually decreases as the impact of frustration due to pinning also gradually diminishes. However, the population that synchronizes first espouses quite distinct routes of the phase transition for different values of $f$, the transition route from ES with no hysteresis ($f=0.99$) turns to second-order ($f=0.9$), and then again ES but with hysteresis ($f=0.8$). Moreover, as the fraction is further decreased to $f=0.7$, both the populations subsequently adopt ES route during forward transition, and simultaneously adopt a second-order route during backward transition. %Also as the fraction is gradually decreased beyond $f=0.99$, the chimeric pattern begins to lose as a complete synchronous population gradually move towards partial synchronous state.

\subsection{Examples of multilayer networks with need for random interlayer pinning model}

Strogatz initiated the idea of random pinning for coupled Kuramoto oscillators in 1989 \cite{strogatz1989}, demonstrating pinning led emerging phenomena. Here, the pinning term drives an individual oscillator to stick to a random phase. %Subsequently, self-feedback random pinning applied to a few driver nodes was utilized for controlling purposes as well~\cite{pinning_control}. 
For multilayer networks, random interlayer accounts for the lack of information about a definite impact of the activities of nodes in one network on those of the other networks.
Such a scheme is more relevant when how activities of one layer get affected by those of other layers is not known. In other words, while the existence of different types of relationships among the same set of interacting units can be clearly evident, making the multilayer model more apt, modeling the nature and structure of multiplexing remains an elusive and non-trivial problem. We attempt to explain this using two examples coming from two completely different systems.

\paragraph{Glial-Neural Multilayer Networks} There have been persistent attempts to understand information processing in the brain under a multilayer network framework consisting of Glial and Neural cells~\cite{Zaikin2021_FrontCell} forming different layers. While modeling intralayer interactions for an individual layer is straightforward, for instance, Glial cells interact with other Glial cells through diffusion, and neural cells interact with each other via synaptic connections. There is not enough knowledge of the nature of Glial-neural interactions for modeling "multiplexing" impact~\cite{Makovkin2017}.
 Another such example is of social multilayer network constructed by different layers representing different social relationships~\cite{Kivela2014}. Here also, the nature of intralayer interactions is rather easy to construe from the available data and hence more easily to model; having such information for multiplexing is non-trivial.

\paragraph{Transport Multilayer Network}
Another such example is of transport systems with their units (stations) connected with each other with transport routes and a mode defining the corresponding layer. There are definite transport networks based on direct connections between different stations for each mode of transport, for instance, flight and train. Additionally, based on the flight or train frequency, the number of airlines operating, inflow, outflow of the passengers at a particular station and other parameters, one can qualitatively model pattern and the nature of pair-wise interactions rather quite accurately. However, similar modeling for multiplexing impact is tricky where one knows for sure that there is an impact of dynamical activities of units and structural patterns of one layer on dynamical activities of units of another layer~\cite{MN_transport2015}.